\documentclass[prl,reprint,superscriptaddress]{revtex4-1}
\usepackage{hyperref}
\usepackage{amsmath}
\usepackage{textcomp}
\usepackage{siunitx}
\usepackage{graphicx}
\usepackage{multirow}
\usepackage{dcolumn}
\newcommand{\SIadj}[2]{\SI[number-unit-product={\text{-}}]{#1}{#2}}

\begin{document}
\title{Sub-ppb measurement of a fundamental band rovibrational transition in HD}
\date{\today}
\author{Arthur Fast}
\affiliation{Max Planck Institute for Biophysical Chemistry, Am Fassberg 11, 37077 G\"ottingen, Germany}
\author{Samuel A. Meek}
\email{samuel.meek@mpibpc.mpg.de}
\affiliation{Max Planck Institute for Biophysical Chemistry, Am Fassberg 11, 37077 G\"ottingen, Germany}

\begin{abstract}
We report a direct measurement of the 0--1 R(0) vibrational transition frequency in ground-state hydrogen deuteride (HD) using infrared-ultraviolet double resonance spectroscopy in a molecular beam.  Ground-state molecules are vibrationally excited using a frequency comb referenced continuous-wave infrared laser, and the excited molecules are detected via state-selective ionization with a pulsed ultraviolet laser.  We determine an absolute transition frequency of \SI{111448815477(13)}{kHz}. The 0.12 parts-per-billion (ppb) uncertainty is limited primarily by the residual first-order Doppler shift.
\end{abstract}

\maketitle

Precise measurements of vibrational transition frequencies in the isotopologues of molecular hydrogen can provide a sensitive probe of fundamental physics. 
Because these transitions can be predicted with high precision using \emph{ab-initio} theory, comparisons between theory and experiment can be used to test quantum electrodynamics, search for new forces beyond the standard model, and determine the proton-electron and deuteron-electron mass ratios more precisely \cite{komasa19,salumbides14,salumbides15,tao18}.
In recent years, many precise measurements of molecular hydrogen transition frequencies have been published \cite{maddaloni10,campargue12,cheng12,kassi12,niu14,mondelain16,wcislo18}, with some recent works reporting fractional uncertainties of less than one part-per-billion (ppb, $10^{-9}$) on vibrational overtone and electronic transition frequencies \cite{altmann18,tao18,cozijn18,fasci18,zaborowski20}.
Many of these recent experiments (with a few notable exceptions \cite{niu14,altmann18}) detect infrared absorption by hydrogen in a gas cell.
In order to determine accurate transition frequencies from such measurements, the data analysis must properly account for the effect of collisions on the line shape and position \cite{wcislo18}.
If saturation techniques are used to achieve sub-Doppler resolution, additional difficulties can arise due to the complex structure of the saturation features \cite{diouf19}.

In this work, we demonstrate a technique that avoids both of these issues by measuring the molecules in the low-density, cold environment of a supersonic molecular beam.
Ground-state hydrogen deuteride (HD) molecules in the beam are vibrationally excited using a tunable continuous-wave (cw) narrow-linewidth infrared (IR) laser referenced to an optical frequency comb (OFC) for absolute accuracy.  
To detect the excitation efficiency, the excited molecules are state-selectively ionized using a pulsed ultraviolet (UV) laser, and the HD$^+$ ions are mass-selectively detected using a time-of-flight mass spectrometer.
This detection scheme is both efficient and nearly background free, making it sensitive enough to detect weak transitions in a sparse sample.
Based on the measured infrared spectra, we are able to determine the absolute frequency of the 0--1 R(0) transition with an uncertainty of \SI{13}{kHz} or 0.12 ppb fractional uncertainty.

\begin{figure}
\includegraphics{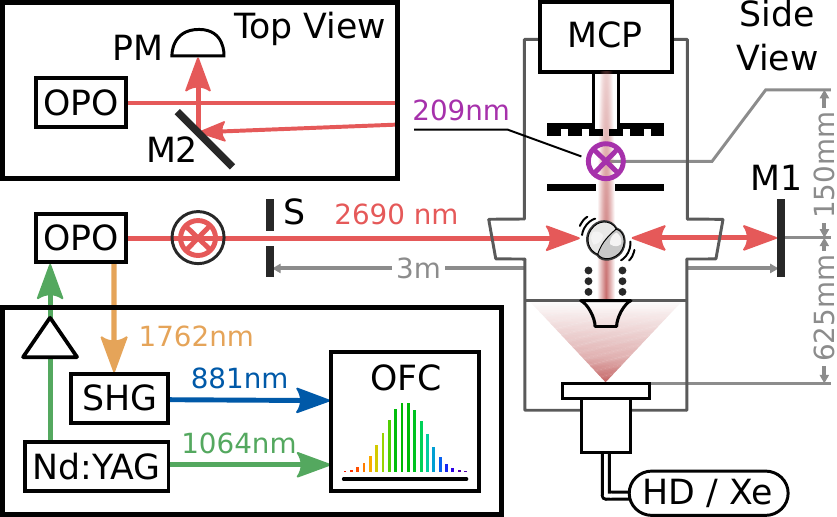}
\caption{Schematic overview of the experimental apparatus.  The main panel shows a side view of the molecular beam.  HD molecules traveling upward are first excited by a \SIadj{2690}{nm} IR spectroscopy laser and are subsequently ionized in a time-of-flight mass spectrometer using a pulsed \SIadj{209}{nm} UV laser.  To compensate for first-order Doppler shifts, the spectroscopy laser is retroreflected from mirror ``M1'' and returns through a slot ``S''.  The top left panel (``Top View'') shows a perpendicular view of outgoing and returning infrared beams near the OPO.  The returning beam is slightly offset so that it can be separated from the outgoing beam using the mirror ``M2'' and detected with a power meter ``PM''. The lower left panel shows the IR laser stabilization scheme.}
\label{fig:setup}
\end{figure}
The infrared spectroscopy laser is produced as the idler of a cw optical parametric oscillator (OPO) based on the design described by Ricciardi et al.\ \cite{ricciardi12}.
The OPO uses a periodically-poled lithium niobate (PPLN) crystal in a bowtie cavity to convert a $\sim$\SIadj{10}{W}, \SIadj{1064}{nm} pump laser into a signal beam at \SI{1762}{nm} and an idler beam at \SI{2690}{nm}.
To measure and stabilize the idler frequency, we use a Ti:Sapphire-based optically-locked OFC, which has been described in detail elsewhere \cite{fast18}.
This comb is stabilized in such a way that the carrier-envelope offset frequency $f_0$ is zero and the mode number $n_p$ has a fixed \SIadj{100}{MHz} offset from a \SIadj{1064}{nm} reference laser.
The OPO pump beam is generated by amplifying the reference laser and thus has the same frequency, $\nu_p$.
The frequency of the signal beam $\nu_s$ is measured by frequency doubling the signal output to \SIadj{881}{nm} using an external PPLN crystal and measuring the frequency of its beat note with the OFC, $f_{\rm{bn},881}$.
Using this beat-note frequency, we can determine the absolute frequency of the idler beam, $\nu_i = \nu_p - \nu_s$, using
\begin{equation}
\nu_i = (n_p - \frac{n_{2s}}{2}) f_r + \SI{100}{MHz} - \frac{f_{\rm{bn},881}}{2}\rm{.}
\end{equation}
In the current measurements, the mode number of the pump beat note $n_p$ is $\num{281631}$,
and the mode number of the signal second-harmonic beat note $n_{2s}$ is \num{340364}.
The comb repetition rate $f_r \approx \SI{999996455.5}{Hz}$ is monitored during each measurement relative to a rubidium oscillator disciplined by a global navigation satellite system (GNSS) receiver.
The lower left inset of Figure \ref{fig:setup} illustrates this scheme for measuring the idler frequency.  

The main body of Figure \ref{fig:setup} shows the molecular beam apparatus.
A mixture of 13\% HD in xenon is expanded upward through a piezo-actuated pulsed valve at a repetition rate of \SI{50}{Hz}.
The molecules pass into a second differentially-pumped chamber and are collimated by a series of rings, resulting in a beam with an angular spread of \SI{3.4}{mrad}.
Approximately \SI{625}{mm} from the valve, the molecules pass through the \SIadj{8}{mm} wide infrared spectroscopy laser, and \SI{150}{mm} further downstream, HD molecules in the $X\, ^1\Sigma^+, v=1, J=1$ state are ionized at the entrance of a time-of-flight mass spectrometer with a pulsed UV laser using 2+1 resonance-enhanced multiphoton ionization (REMPI) through the $EF\, ^1\Sigma^+, v=0, J=1$ intermediate state.
The UV laser is a frequency-tripled pulsed dye laser that produces \SI{\sim 1}{mJ} per pulse at \SI{209}{nm}, and the pulse arrives \SI{1.95}{ms} after the molecules leave the nozzle, selecting HD molecules with a velocity of \SI{400}{m/s}.
The extraction field in the mass spectrometer is switched on during a $\pm$\SI{100}{\us} window around the ionization pulse but is switched off while the molecules are passing through the infrared laser to minimize fields in the spectroscopy region.

\begin{figure}
\includegraphics{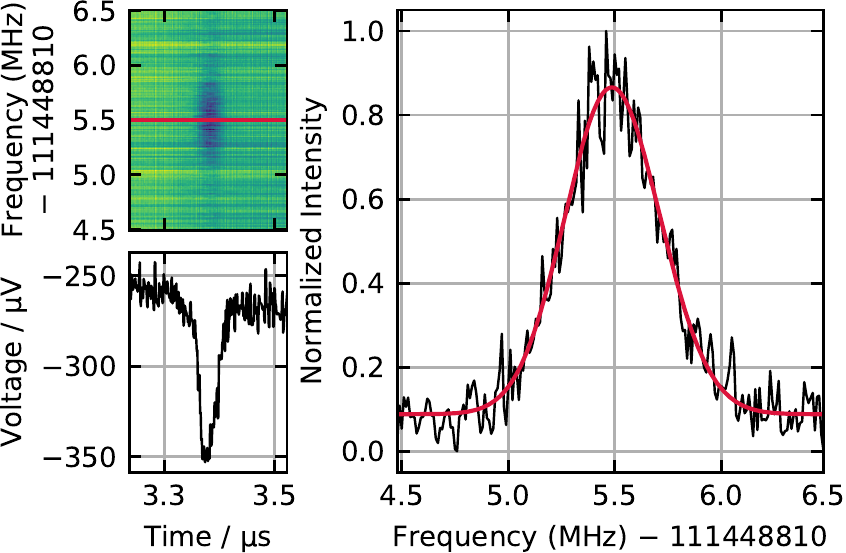}
\caption{Data for a typical infrared spectrum.  The upper left panel shows the ion signal as a function of time delay after the ionization laser pulse and infrared laser frequency, while the lower left panel shows a cut through of this data at a single infrared laser frequency.  The infrared spectrum given by the total HD$^+$ signal at each frequency is shown in the right panel in black, while the red curve shows a multi-Gaussian fit.}
\label{fig:avgmatrix}
\end{figure}
Spectra are measured by recording the time-resolved ion signal while scanning the frequency of the infrared laser.
The upper left panel of Figure \ref{fig:avgmatrix} shows a typical averaged measurement of the ion signal as a function of time delay after the laser pulse and infrared laser frequency, while the lower panel shows the ion signal versus time delay for a single laser frequency.
The Gaussian peak at \SI{3.38}{\us} corresponds to the HD$^+$ mass channel.
To determine the total ion intensity in this peak, the peak center and its standard deviation $\sigma$ are first computed by averaging time traces at all laser frequencies and fitting the trace with the sum of a Gaussian function and a linear background.  
The ion signal is then computed at each laser frequency by averaging the signal over a $\pm 2 \sigma$ region around the peak and subtracting a background calculated by averaging over regions covering $(-8\sigma$,$-4\sigma)$ and $(+4\sigma$,$+8\sigma)$ relative to the peak.
The black curve in the right panel of Figure \ref{fig:avgmatrix} shows the normalized ion intensity as a function of laser frequency, while the red curve shows a fit to this data using five overlapping Gaussian peaks.

In order to determine the absolute 0--1 R(0) transition frequency from the measured spectra, a number of potential systematic shifts have been considered.
The recoil shift $h \nu_0^2 / (2 m_{\rm HD} c^2) \approx \SI{9.1}{kHz}$ and the second-order Doppler shift $-\nu_0 v^2/(2 c^2) \approx \SI{-0.1}{kHz}$ can be computed with high accuracy and are corrected in the reported value.
Other effects are found to have a negligible influence on the measured transition frequency.
Based on the HD polarizability computed by Ko\l{}os et al.\ \cite{kolos67}, the alternating-current (ac) Stark shift is estimated to be less than \SI{1}{Hz} for the laser intensity used in the experiment.
External Helmholtz coils are used to reduce the magnetic field in the spectroscopy region to below \SI{3}{\micro \tesla}, resulting in a residual Zeeman shift of less than \SI{100}{Hz}.
The gas density in the spectroscopy region is estimated to be $\sim 3\times10^{17}$ molecules per cubic meter; applying the \SI{-10}{kHz/Pa} pressure shift reported by Cozijn et al.\ \cite{cozijn18} for the 0--2 R(1) transition would result in an estimated pressure shift of \SI{10}{Hz}.
Errors in the frequency of the rubidium reference are expected to contribute less than \SI{0.5}{kHz} to the overall uncertainty.

Two systematic effects are not so easily ignored and must be considered in further detail.
The first and most significant is the residual shift due to the first-order Doppler effect.
Although the infrared laser is nominally aligned so that its propagation direction is perpendicular to the central velocity of the molecular beam, even a small error in this alignment can result in a significant shift of the measured transition frequency.  
To detect such a shift, the infrared laser is retroreflected after passing through the spectroscopy region and interacts with the molecules a second time.
If the retroreflection were perfect in both direction and amplitude, the Doppler shift of the second beam would be equal and opposite to the first and both beams would contribute equally to the vibrational excitation, resulting in no net shift.
Unfortunately, an angular deviation between the outgoing and returning laser beams along the molecular beam direction or an imbalance of the amplitudes would result in imperfect cancellation.  

To limit the angular deviation between the two beams, a \SIadj{7}{mm} wide slot near the OPO (labeled ``S'' in Figure \ref{fig:setup}) \SI{\sim 3}{m} from the retroreflection mirror (``M1'') helps constrain the offset between the outgoing and returning beams along the molecular beam direction.
The slot is aligned so that it is centered vertically on the outgoing beam, and the returning beam must pass through the same slot to reach a power meter (``PM'' in Figure \ref{fig:setup}).
The returning beam is offset by about \SI{7}{mm} from the outgoing beam in the direction perpendicular to the molecular beam (see the ``Top View'' panel of Figure \ref{fig:setup}) to facilitate the power measurement.
Based on the sensitivity of the measured power to changes of the slot height, we estimate an uncertainty of the offset between the outgoing and returning beams along the molecular beam direction of \SI{0.5}{mm}, which translates to a \SI{12}{kHz} uncertainty of the infrared transition frequency.
This error is included both as a random error that contributes to the uncertainty of the transition frequency extracted from each spectrum and as a potential systematic error that shifts all spectra in the same direction.  

\begin{figure}
\includegraphics{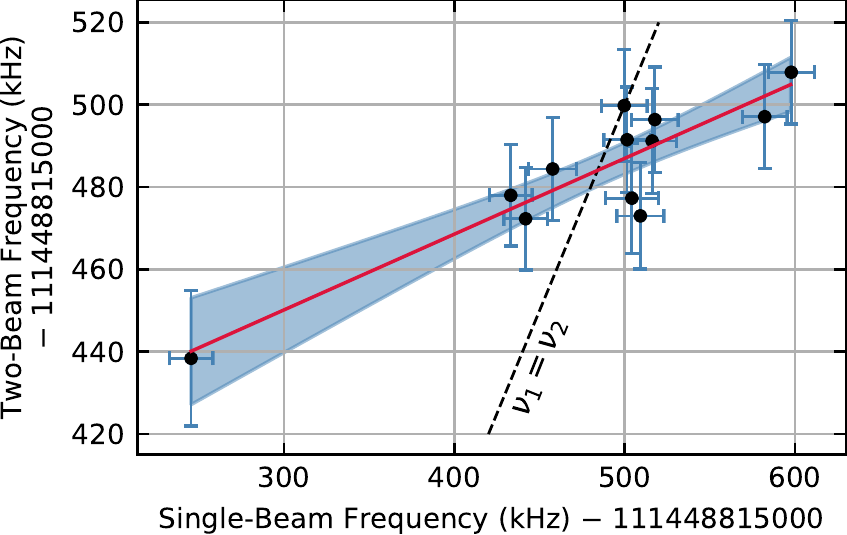}
\caption{Transition frequency pairs extracted from spectra measured with a retroreflected spectroscopy laser (``Two-Beam Frequency'') and with a single pass of the spectroscopy laser (``Single-Beam Frequency'') in close succession.  The red line shows an orthogonal distance regression fit of the data, with a $1\sigma$ confidence region indicated in blue. The Doppler-corrected transition frequency is given by the crossing between the linear fit and the (dashed) $\nu_1 = \nu_2$ line.}
\label{fig:odrfit}
\end{figure}
The mismatch in amplitude between the two beams (caused by losses in the window and retroreflection mirror) is compensated by measuring each spectrum both with and without the retroreflected beam.  
If the transition frequency determined with both beams ($\nu_2$) is the same as the frequency determined with one beam ($\nu_1$), then the laser is perpendicular to the molecular beam, but a difference between the two frequencies indicates that $\nu_2$ has been shifted from its true value by an amount proportional to $\nu_1 - \nu_2$.
To apply this concept to the measured data, we fit all twelve measured frequency pairs $(\nu_1,\nu_2)$ with a linear model $\nu_2 = a \nu_1 + b$ using an orthogonal distance regression fit \cite{boggs89} in order to account for the uncertainties in both coordinates.
The Doppler-corrected frequency is determined by finding the crossing point between the linear model and the line $\nu_1 = \nu_2$, which occurs at $\nu_1 = \nu_2 = b/(1-a)$.
The uncertainty of this crossing point is determined by propagating the errors given by the fit covariance matrix for $a$ and $b$.  
Figure \ref{fig:odrfit} illustrates the results of such a fit.

\begin{figure}
\includegraphics{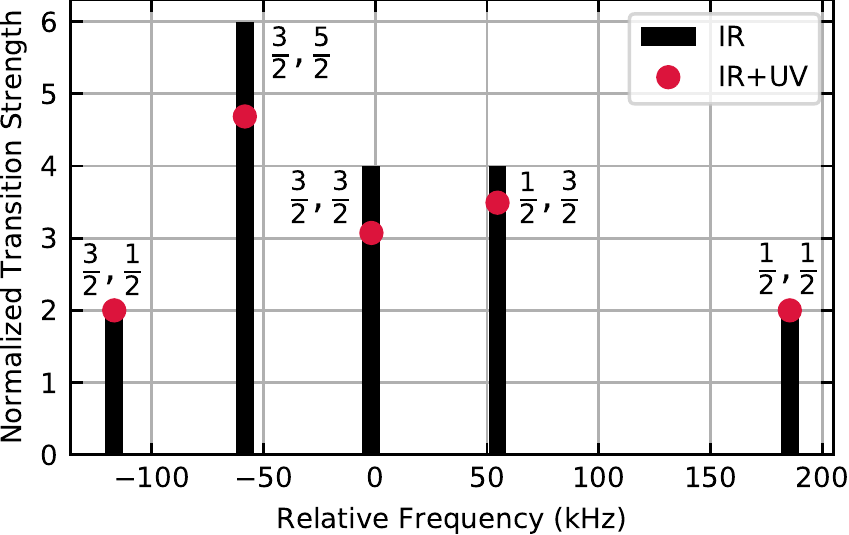}
\caption{Hyperfine substructure of the R(0) transition.  The horizontal positions of the black bars show the frequencies of the hyperfine-resolved vibrational transitions relative to the line center, while their heights show the relative IR transitions intensities.  The fractional values next to each peak indicate the $F_1$ and $F$ quantum numbers of the $v'=1,J'=1$ excited state.  Red dots show the relative efficiencies for IR vibrational excitation followed by two-photon UV electronic excitation.}  
\label{fig:hyperfine}
\end{figure}
Uncertainties in the relative contributions of the individual hyperfine components in the transition can also contribute an error to the measured transition frequency.  
The R(0) transition contains nine hyperfine components which, due to the degeneracy between the $F=1/2$ and $F=3/2$ levels in the ground state, results in five unique transition frequencies spread over \SI{\sim 300}{kHz}.
Figure \ref{fig:hyperfine} shows the positions of these components relative to the transition center.
These individual components are not resolved in our experiment but are blended into a single \SIadj{\sim 500}{kHz} wide peak.
It is therefore important to accurately predict the relative intensities of these five components in order to correctly determine the transition center from a measured spectrum.

To model the hyperfine substructure, we use an effective Hamiltonian defined by
\begin{equation}
\begin{aligned}
\hat{H} =& T_v + B_v \hat{N}^2 - D_v \hat{N}^4 +
	c_{H,v} \hat{I}_H \cdot \hat{N} + c_{D,v} \hat{I}_D \cdot \hat{N} +\\
	& \frac{e Q q_{0,v}}{4 I_D (2 I_D - 1)} \sqrt{6} T^2_{q=0} (\hat{I}_D, \hat{I}_D)
	- S_v \sqrt{6} T^2_{q=0} (\hat{I}_H, \hat{I}_D)\rm{.}
\end{aligned}
\end{equation}
A separate set of parameters is used for each vibrational state.
The operator $\hat{N}$ is the rotational angular momentum and $\hat{I}_H$ and $\hat{I}_D$ are the spins of the hydrogen and deuterium nuclei, respectively.
Band origins $T_v$ and rotational constants $B_v$ and $D_v$ are determined by fitting the energies calculated by the program H2Spectre \cite{komasa19} for the first three rotational levels in each vibrational state.
The hyperfine parameters $c_{H,v}$, $c_{D,v}$, $e Q q_{0,v}$, and $S_v$ (``$c_{\rm dip}$'') are taken from Dupr\'e \cite{dupre20}; the signs of $c_{H,v}$ and $c_{D,v}$ have been inverted to correctly reproduce the results from that work.
Eigenenergies and eigenvectors based on these parameters are calculated using the program spcat \cite{pickett91}.

We then define $P_{\rm IR}$ as the normalized sum of one-photon transition strengths from any of the degenerate $X\, ^1\Sigma^+, v''=0, J''=0$ hyperfine levels to a specific $M_F'$ level in the $X\, ^1\Sigma^+, v'=1, J'=1$ state due to an IR laser polarized along the $Z$ axis.  
\begin{equation}
\begin{aligned}
&P_{\rm IR} (F_1', F', M_F') = A_{\rm IR} \\
&\sum_{F'',M_F''} |\langle v'',J'',F_1'',F'',M_F''|\hat{\mu}_Z|v',J',F_1',F',M_F'\rangle|^2
\end{aligned}
\label{eqn:PIR}
\end{equation}
In general, $F_1$ (defined by $\hat{F}_1 = \hat{N} + \hat{I}_H$) is not a good quantum number; the symbol $|v',J',F_1',F',M_F'\rangle$ is used here as a shorthand for the eigenstate with the largest contribution from the corresponding basis vector.  
Because the linewidth of the UV laser is broad enough to cover all hyperfine components, the ionization efficiency from a particular $M_F'$ level in the $X\, ^1\Sigma^+,v'=1,J'=1$ state is modeled as the normalized sum of two-photon transition strengths to any $EF\, ^1\Sigma^+,v=0,J=1$ level due to a UV laser polarized along the $X$ axis.
\begin{equation}
\begin{aligned}
&P_{\rm UV} (F_1', F', M_F') = A_{\rm UV} \\
&\sum_{F_1,F,M_F} |\langle X,v',J',F_1',F',M_F'|\hat{\mu}_X^2|EF,v,J,F_1,F,M_F\rangle|^2
\end{aligned}
\label{eqn:PUV}
\end{equation}
The normalization factors $A_{\rm IR}$ and $A_{\rm UV}$ are chosen such that the average values of $P_{\rm IR}$ and $P_{\rm UV}$ are 1.

The amplitudes of the black bars in Figure \ref{fig:hyperfine} show the strengths of the hyperfine components of the transition calculated by summing $P_{\rm IR} (F_1', F', M_F')$ over all $M_F'$.
If the detection efficiency of the UV laser is taken into account by instead summing $P_{\rm IR} (F_1', F', M_F') \times P_{\rm UV} (F_1', F', M_F')$ over all $M_F'$, it is found that certain transitions are detected less efficiently, as indicated by the red dots.  
The intensities predicted by the second model (red dots) only hold if there is no saturation of the UV transition and no reorientation of the molecules between the IR excitation and UV ionization lasers.
If either of these conditions does not hold, the relative intensities of the hyperfine components will be more closely described by the first model (black bars).  
To account for this possibility, we analyze the measured spectra using the intensities predicted by both models and report the average of the two results as a best estimate; half of the difference is then included in the error budget.  
The Doppler-corrected peak position is determined to be \SI{111448815.4871(47)}{MHz} using the first model (IR only) and \SI{111448815.4840(47)}{MHz} using the second (IR and UV), resulting in an average of \SI{111448815.4856(47)}{MHz}.

Table \ref{tab:contributions} summarizes the contributions to the measured transition frequency.
After correcting for shifts due to recoil and second-order Doppler effects, we conclude an absolute frequency for the HD 0--1 R(0) transition of \SI{111448815.477(13)}{MHz}, with the uncertainty dominated by residual first-order Doppler shifts.
Table \ref{tab:comparison} shows a comparison between this result and previous theoretical and experimental values.
The experimental value is determined by combining the 0--1 Q(1) transition frequency reported by Niu et al.\ \cite{niu14} with the $J=0 \rightarrow 1$ rotational transition frequency reported by Drouin et al.\ \cite{drouin11}; the theoretical value is computed using H2Spectre \cite{komasa19}.
The present result agrees with both previous values but shows a factor of 500 smaller uncertainty than the experimental result and 50 smaller than the theoretical.
Interestingly, we note that five other measurements of hydrogen vibrational transition frequencies \cite{cheng12,mondelain16,wcislo18,diouf19,zaborowski20}, covering all three stable isotopologues, show fractional deviations from the theoretical predictions from H2Spectre consistent with the $8.7\times10^{-9}$ deviation measured here to within experimental uncertainty.

\begin{table}
\caption{Frequency and uncertainty contributions to the determination of the 0--1 R(0) transition frequency.}
\begin{ruledtabular}
\begin{tabular}{lll}
Contribution&	Frequency [kHz]&	$\sigma$ [kHz]\\
\hline
Doppler-corrected peak position&	\num{111448815485.6}&	4.7\\
First-order Doppler&	0&	12\\
Hyperfine&	0&	1.6\\
Frequency reference error&	0& $<$0.5\\
ac Stark, Zeeman, Pressure&	0&	 $<$0.1\\
Recoil&	-9.1& 0\\
Second-order Doppler& +0.1& 0\\
Total&	\num{111448815477}&	13\\
\end{tabular}
\end{ruledtabular}
\label{tab:contributions}
\end{table}

\begin{table}
\caption{Comparison between the HD 0--1 R(0) transition frequency determined in the present work and previous experimental and theoretical values.}
\begin{ruledtabular}
\begin{tabular}{lll}
&	Frequency [MHz]&	Deviation [MHz]\\
\hline
This work&	\num{111448815.477}(13)&\\
Refs. \cite{niu14} and \cite{drouin11}&	\num{111448818.5}(6.6)&	\num{-3.0}(6.6)\\
Ref. \cite{komasa19}&	\num{111448814.5}(6)&	\num{1.0}(6)\\
\end{tabular}
\end{ruledtabular}
\label{tab:comparison}
\end{table}

The precision of the current result is limited primarily by residual first-order Doppler shifts and possible shifts due to unresolved hyperfine structure.
We anticipate that, by improving the retroreflection quality and characterizing the hyperfine effects by changing the relative polarizations between the two lasers, the uncertainty can be reduced below \SI{1}{kHz}, or $10^{-11}$ fractional uncertainty.
Measurements at this level of precision, combined with accurate theoretical predictions, would result in values for the proton-electron and deuteron-electron mass ratios that are more precise than the 2018 CODATA recommended values \cite{codata18}.
With improved sensitivity, the same techniques used here could also be used to measure quadrupole transitions in the homonuclear isotopologues, making it possible to investigate the properties of the proton and deuteron separately.

\begin{acknowledgments}
We gratefully acknowledge S.\ Kaufmann and K.\ Papendorf for lending us the UV ionization laser, as well as M.\ De Rosa for his extensive advice on the design of the OPO.
\end{acknowledgments}
\bibliography{bib}
\end{document}